\documentclass[runningheads, a4paper]{llncs}
\usepackage
[
        a4paper,% other options: a3paper, a5paper, etc
        left=2cm,
        right=2cm,
        top=1.5cm,
        bottom=1.5cm,
]
{geometry}

\usepackage{geometry}
\usepackage{filecontents}
\usepackage[document]{ragged2e}

\usepackage{graphicx}
\usepackage{authblk}
\usepackage{amsmath}
\usepackage{amsfonts}
\usepackage{amssymb}
\usepackage{tikz}
\usepackage{cite}
\usepackage{float}
\usepackage{filecontents}
\usepackage[utf8]{inputenc}
\usepackage{fancyhdr} % Custom headers and footers
\begin{document}

\title{Azimuthal anisotropy measurement of $\phi$ mesons in Au+Au collisions at $\sqrt{s_{NN}}$ = 27 and 54.4 GeV at STAR}
\author{Prabhupada Dixit (for the STAR Collaboration) }
\institute{Indian Institute of Science Education and Research, Berhampur, India \\prabhupadad@iiserbpr.ac.in}

\maketitle

\begin{abstract}
\fontsize{9}{11}\selectfont
The elliptic flow coefficient ($v_{2}$) of $\phi$ mesons at mid-rapidity  as a function of transverse momentum ($p_{T}$) is measured for Au+Au collisions at  centre of mass energy ($\sqrt{s_{NN}}$) 27 and 54.4 GeV. The $v_{2}$ measurement is done using event plane from Time Projection Chamber (TPC, $|\eta|$ $<$1.0) and Event Plane Detectors (EPD, 2.1$<$$|\eta|$$<$5.1) for 54.4 and 27 GeV, respectively.  A high precision test of the number of constituent quark scaling of $\phi$ meson $v_{2}$ (by including measurements for other hadrons) has been shown. The results are compared to transport model (AMPT) calculations. 
%\keywords{First keyword  \and Second keyword \and Another keyword.}
\end{abstract}
\section{Introduction}
\fontsize{11}{12}\selectfont
The primary objective of relativistic heavy-ion collision is to produce a medium in a small scale that has resemblances to the initial stage of the universe created in the big bang. In this extremely hot and dense medium, the matter exists in a new state where quarks and gluons are no longer bound inside the hadrons. This state is called Quark-Gluon-Plasma (QGP)\cite{SHURYAK198071}. To study the various macro-state properties of QGP, azimuthal anisotropy in its expansion is used as one of the well-known observables\cite{PhysRevLett.78.2309}. Due to the difference in pressure gradient along different axes of the collision system, the expansion of medium is not isotropic in space and this spatial anisotropy gives rise to the anisotropy in momentum space. To study different order of azimuthal anisotropy, Fourier expansion of the azimuthal distribution of particles in momentum space is used\cite{Voloshin:1994mz}, which is given by
\begin{equation}
E\frac{d^{3}N}{dp^{3}} =\frac{1}{2\pi}\frac{d^{2}N}{p_{T}dp_{T}dy}\left[1 + \sum_{n} 2v_{n}\cos n(\phi-\psi_{R})\right].
\end{equation}

The coefficient $v_{n}$ is called $n^{th}$ order flow coefficient which is given by
\begin{equation}
v_{n} = \langle \cos n(\phi -\psi_{R})\rangle.
\end{equation}

The angled bracket indicates the average over all the particles and all the events. $\psi_{R}$ is the reaction plane (plane spanned by the impact parameter vector and collision axis) of the collision system. In this paper, we have studied the second-order flow coefficient $v_{2}$, known as elliptic flow, which is analogous to the eccentricity of the ellipse in the momentum space.\\
The elliptic flow of $\phi$ mesons is considered as a good probe for the initial stage of the medium produced in heavy-ion collisions. Due to its small hadronic interaction cross-section and early freeze-out, $\phi$-meson $v_{2}$ is less affected by the late-stage hadronic interactions\cite{PhysRevLett.54.1122}. 
$\phi $ meson has mass (1.02 $\rm GeV/c^{2}$) comparable to the mass of the lightest baryons (protons) and at the same time it is meson, therefore the elliptic flow of $\phi$ mesons can be used to investigate the effect of mass vs. particle type dependence in $v_{2}$. In this proceedings, elliptic flow of $\phi$ mesons at mid-rapidity as a function of transverse momentum in Au+Au collision at $\sqrt{s_{NN}}$ = 27 and 54.4 GeV is presented. 

\section{STAR Detectors}

The main STAR detectors that we have used in our analysis is the Time Projection Chamber (TPC), Time Of Flight (TOF) detector and Event Plane Detectors (EPDs) \cite{SHAO2006419, ADAMS2020163970}. The TPC is a cylindrical detector having a length of 4.2m and a radius of 4m which surrounds the beam pipe. It has full azimuthal coverage and has pseudorapidity coverage of $|\eta|$$<$1.8. The TPC is used to detect  charged particles by measuring their ionization energy loss (dE/dx). Surrounding the TPC we have TOF which also has full azimuthal coverage and pseudorapidity coverage $|\eta|$$<$1. The TOF is used to identify charged particles with high $p_{T}$. We have two newly installed EPDs located at $\pm$ 3.75 m from the center of the TPC. The EPDs have pseudorapidity coverage of 2.1$<$$|\eta|$$<$5.1. These EPDs are used to construct the event plane of the collision.
%\linenumbers
\section{Flow Analysis Method}
We first reconstruct $\phi$ mesons by calculating the invariant mass of its daughter particles $K^{+}$ and $K^{-}$ from the same events. The combinatorial background is calculated by using the event-mixing method. The normalized background distribution is subtracted from the signal+background distribution from the same events to get the signal of $\phi$ meson. The signal is then fitted with the Breit-Wigner function to calculate the yield.

%\begin{frame}
%\frametitle{}  % no frame title
%\begin{figure}[H]\centering
  %\includegraphics[width=0.30\textwidth]{sig+bag.eps}
  %\includegraphics[width=0.30\textwidth]{signal.eps}
  %\caption[]{The left panel shows the invariant distribution of signal+background and the only combinatorial constructed by event mixing method. The Right panel shows the only signal of phi meson after subtraction of normalized background. The distribution in the right panel is fitted with the Breit-Wigner function (for the peak) and a 1st order polynomial function (for the residual background)}
%\end{figure}
%\end{frame}

Next, we need to calculate the reaction plane of the collision. Since the impact parameter can not be  measured experimentally, the event plane is used as a proxy of the reaction plane. The event plane is calculated using final state particles as follows:
\begin{equation}
\psi_{2} =\frac{1}{2} \tan^{-1}\Big(\frac{\sum_{i} w_{i} \sin(2\phi_{i})}{\sum_{i} w_{i} \cos(2\phi_{i})}\Big),
\end{equation}
where $w_{i}$ is the weight factor equals to $p_{T}$ $\times$ $\phi$-weight in our analysis. Here $\phi$-weight accounts for the azimuthal acceptance correction of the detectors.

We have used the second-order TPC $\eta$ sub-event plane for 54.4 GeV and EPD full event plane for 27 GeV to calculate $v_{2}$. Since the number of particles produced in an event is finite, the observed $v_{2}$ needs to be corrected by the event plane resolution. The final corrected $v_{2}$ is given by

 \begin{equation}
v_{2}= \frac{\langle \cos 2(\phi -\psi_{2})\rangle}{\sqrt{\langle \cos 2(\psi_{A}- \psi_{B})\rangle}} .
 \end{equation}

 The denominator of the above equation is called resolution. $\psi_A $and $\psi_{B}$ are the second order sub-event planes constructed from the particles within -1$<$$\eta$$<$0 and 0$<$$\eta$$<$1 respectively. 
 The resolution given in the above equation is for the sub-event plane. The full-event plane resolution can be calculated from the sub-event plane resolution as described in this reference\cite{PhysRevC.58.1671}.
 The resolution as a function of centrality is shown in Fig. 1.
 \begin{figure}[H]\centering
\includegraphics[scale=.37]{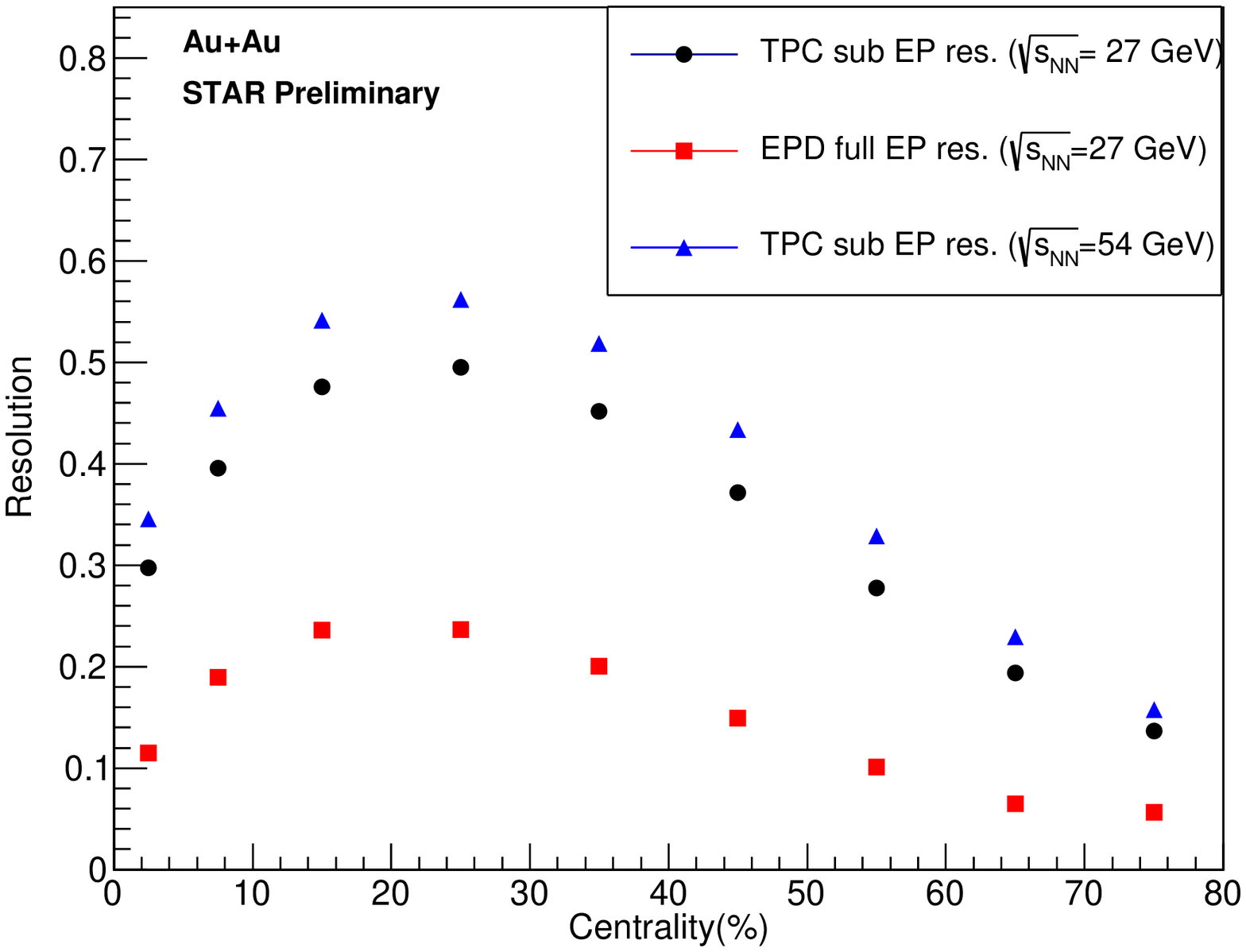}
\includegraphics[scale=.34]{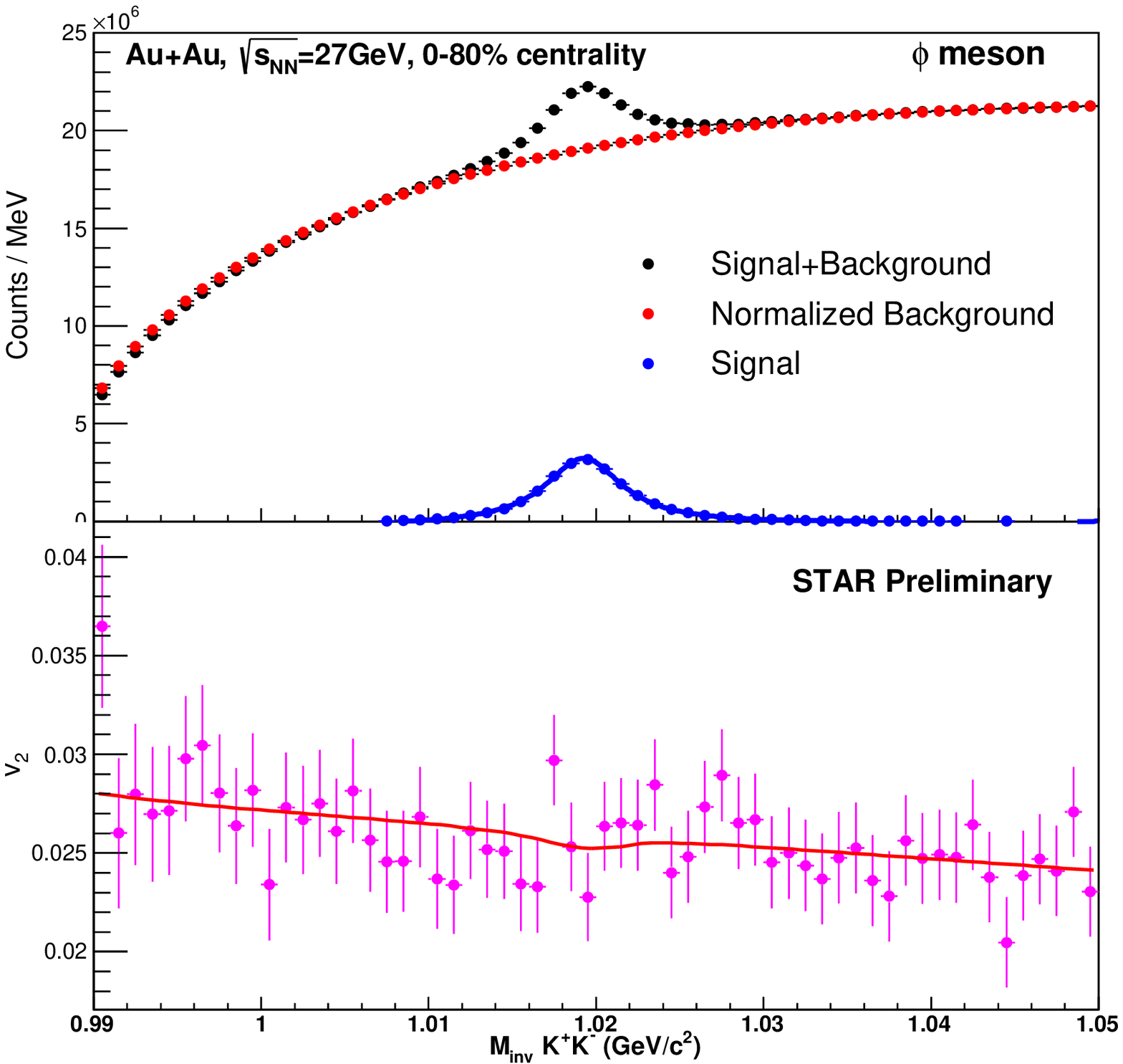}
\caption[]{Left panel shows TPC sub-event plane resolutions for $\sqrt{s_{NN}}$ = 27 and 54.4 GeV as a function of centrality. EPD full-event plane resolution for  $\sqrt{s_{NN}}$ = 27 GeV is also shown. In the upper right panel, invariant mass distribution of  $K^{+}$ and $K^{-}$ from the same events (signal + background), from the mixed events (normalized background) and the background subtracted signal is shown. In the lower right panel,  $v_{2}$ using EPD event plane is plotted as a function of the invariant mass of $K^{+}$ and $K^{-}$. The distribution is fitted with Eq. (5).}
\end{figure}
 $v_{2}$ of $\phi$ mesons has been calculated by using the invariant mass method\cite{PhysRevC.70.064905}, in which, $v_{2}$ is calculated as a function of the invariant mass of $K^{+}$ and $K^{-}$ as shown in the lower panel of Fig. 1 (right) and the distribution is fitted with Eq. (5).
\begin{equation}
v_{2}^{S+B} (M_{inv})= v_{2}^{S} \frac{S}{S+B}(M_{inv}) + v_{2}^{B}\frac{B}{S+B}(M_{inv})
\end{equation}
where $v_{2}^{S}$ is  $v_{2}$ of signal and  $v_{2}^{B}$ is  $v_{2}$ of background. S is the yield of the $\phi$ meson and B accounts for the total background counts. $v_{2}^{B}$ can be approximated as a first-order polynomial function of invariant mass and $v_{2}^{S}$ can be obtained as a free parameter of the fit. \par
%The high precision measurement of $\phi$ mesons $v_{2}$ is expected to play an important role to study the collective expansion of the QGP medium. 
In 2017 and 2018, the STAR experiment has collected a large amount of Au+Au data at 54.4 and 27 GeV, respectively. We have analyzed $v_{2}$ of $\phi$ mesons using 600 M (54.4 GeV) and 350 M (27 GeV) Au+Au minimum bias events. In the year 2018, 27 GeV Au+Au data were collected using EPD installed in the STAR detector system.
 
 \section{Results}
 \subsection{$p_{T}$ dependence of $v_{2}$}

 The $p_{T}$ dependence of $\phi$ meson $v_{2}$ has been studied at $\sqrt{s_{NN}}$ = 27 GeV using EPD event plane and compared to  $v_{2}$ using TPC event plane\cite{PhysRevC.88.014902} for 0-80$\%$ centrality events. TPC and EPDs have different $\eta$ coverage and have different minimum $\eta$ gap between reconstructed $\phi$ mesons and the event plane (0.05 for TPC and 1.1 for EPD) but as shown in Fig. 2 (left panel) the $v_{2}$ for both the event planes are compatible with each other within statistical uncertainties which indicates that the effect of non-flow is negligible.
 In the right panel, we have compared the high precision measurement of $\phi$ meson $v_{2}$ at $\sqrt{s_{NN}}$ = 54.4 with that at 62.4 GeV and found that they are compatible with each other. 
 \begin{figure}[H]\centering
  \includegraphics[width=0.45\textwidth]{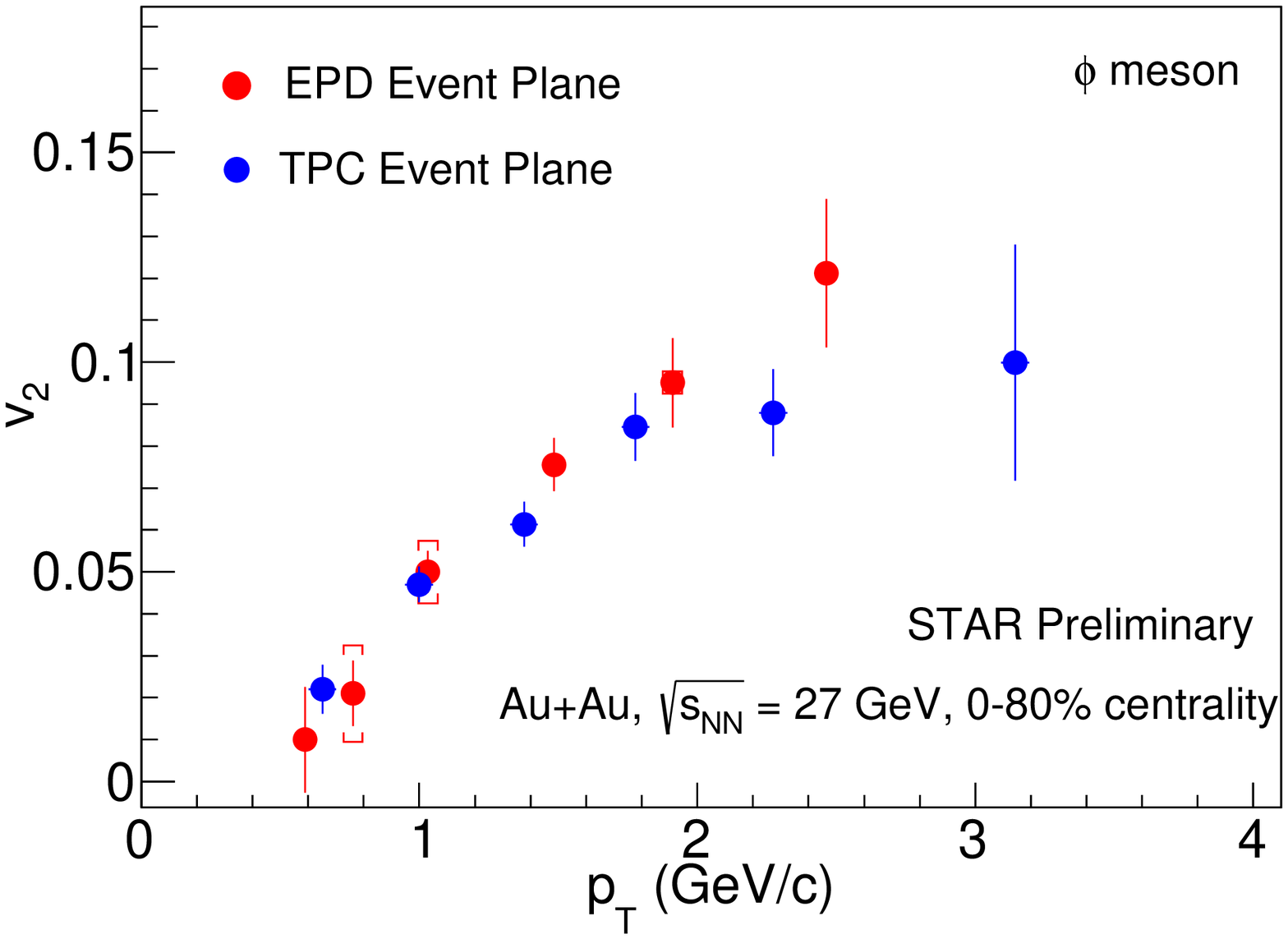}
  \includegraphics[width=0.46\textwidth]{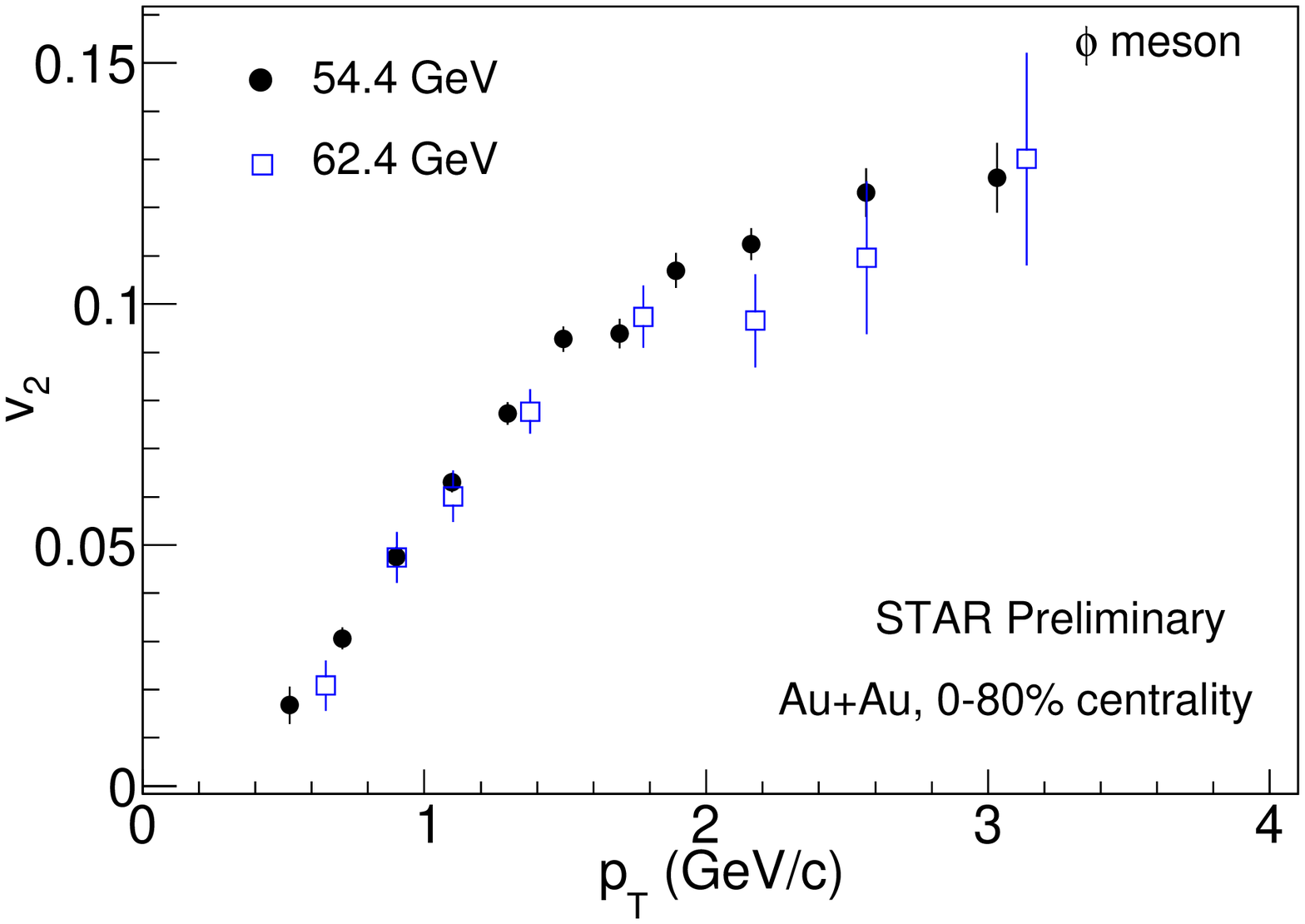}
  \caption[]{Left panel shows the comparison of $v_{2}$ using TPC event plane and EPD event plane at $\sqrt{s_{NN}}$ =27 GeV for centrality 0-80$\%$ and the right panel shows the comparision between  $v_{2}$ at $\sqrt{s_{NN}}$ =54.4 GeV and 64.4 GeV for centrality 0-80$\%$. Vertical lines represent statistical uncertainties and cap-symbols represent systematic uncertainties. }
\end{figure}
\subsection{NCQ scaling}
One of the main goals of the STAR experiment is to search for turn-off signal of QGP which can be studied by the number of constituent quark (NCQ) scaling in $v_{2}$ for all hadrons as a function of energy. The number of constituent quark scaling has been studied at $\sqrt{s_{NN}}$ = 27 and 54.4 GeV using $v_{2}$ of $\phi$ as well as other identified hadrons\cite{PhysRevC.88.014902}. The results are shown in Fig. 3.
\begin{figure}[H]\centering
\includegraphics[scale=0.48]{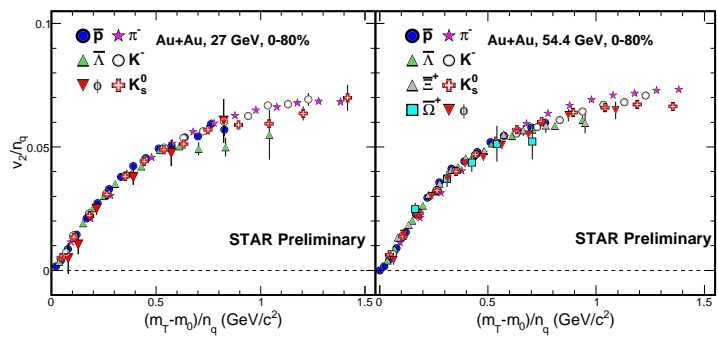}
\caption[]{ $v_{2}$ scaled by the number of constituent quarks as a function of transverse kinetic energy per number of constituent quarks for various charged hadrons are shown at                                $\sqrt{s_{NN}}$ = 27 and 54.4 GeV for 0-80$\%$ centrality. Only statistical error bars are shown in Fig.3.}
\end{figure}
The NCQ scaled $v_{2}$ is found to be similar for all the particles under study.
This observation can be interpreted as the collectivity being developed at the partonic stage of the evolution of the system in Au+Au collision at $\sqrt{s_{NN}}$ = 27 and 54.4 GeV.

\subsection {The AMPT model comparison}
The measured elliptic flow of $\phi$ mesons is compared to  A Multi-Phase Transport (AMPT) string melting model\cite{PhysRevC.72.064901} as shown in Fig. 4. The AMPT uses the same initial conditions as in HIJING. Scattering among partons is modeled by Zhang’s parton cascade. The interactions between partons in the AMPT model give rise to $v_{2}$. The parton-parton interaction cross-section is taken to be 3 mb. Figure 4 shows that the AMPT model describes the data for $\sqrt{s_{NN}}$ = 54.4 GeV up to $p_{T}$ = 2.0 GeV/c, whereas the model overpredicts the data for 27 GeV.
 \begin{figure}[H]\centering
  \includegraphics[width=0.80\textwidth]{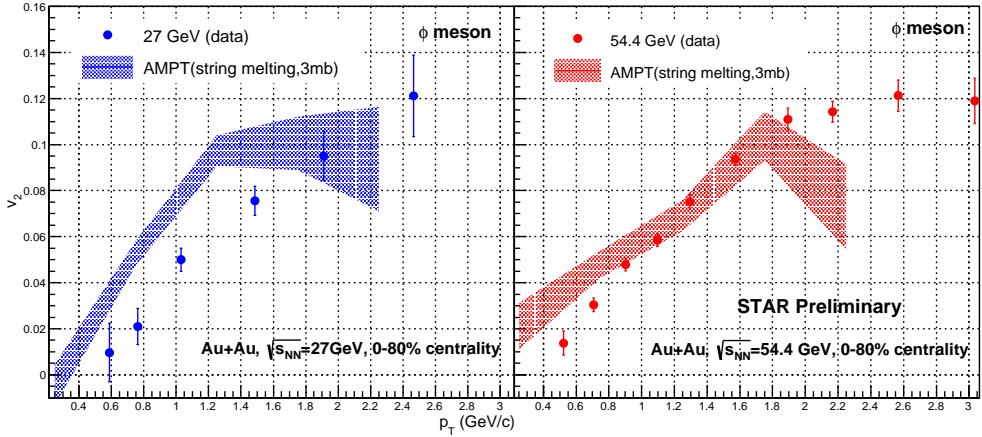}
  \caption[]{The left panel shows the comparison of $p_{T}$ dependence of $v_{2}$ for 27 GeV data with string melting AMPT model with parton-parton cross-section 3 mb. The right panel shows the same for 54.4 GeV data. }
\end{figure}

\section {Summary}
The elliptic flow coefficient $v_{2}$ of $\phi$ mesons at $\sqrt{s_{NN}}$ = 27 and 54.4 GeV has been studied for Au+Au collision. $v_{2}$ of $\phi$ meson at $\sqrt{s_{NN}}$ = 27 GeV using EPD event plane is found to be in good agreement with $v_{2}$ using TPC event plane. The NCQ scaling is observed for all the hadrons at both these energies which indicates the partonic collectivity developed inside the medium during the earlier stage of its evolution. The AMPT with string melting model well predicts the data for 54.4 GeV than those for 27 GeV.

\bibliographystyle{unsrt}%\bibliographystyle{apacite}

\bibliography{ref}
\end{document}